\documentclass[sigconf]{acmart} 
\AtBeginDocument{
  }
\copyrightyear{2025}
\acmYear{2025}
\setcopyright{cc}
\setcctype{by}
\acmConference[CHI '25]{CHI Conference on Human Factors in Computing Systems}{April 26-May 1, 2025}{Yokohama, Japan}
\acmBooktitle{CHI Conference on Human Factors in Computing Systems (CHI '25), April 26-May 1, 2025, Yokohama, Japan}\acmDOI{10.1145/3706598.3713279}
\acmISBN{979-8-4007-1394-1/25/04}

\usepackage{enumitem}
\usepackage{longtable}
\usepackage{booktabs}
\usepackage{tabularx}

\AtBeginDocument{
  }

\newcommand{\add}[1]{\textcolor{black}{#1}}

\begin{document}

\title[How Millions of Fans Build Collective Understanding of Algorithms and Organize Coordinated Algorithmic Actions]{Let's Influence Algorithms Together: How Millions of Fans Build Collective Understanding of Algorithms and Organize Coordinated Algorithmic Actions}

\author{Qing Xiao}
\affiliation{
  \institution{Human-Computer Interaction Institute, Carnegie Mellon University}
  \city{Pittsburgh}
  \state{Pennsylvania}
  \country{USA}
}
\email{qingx@cs.cmu.edu}

\author{Yuhang Zheng}
\affiliation{
  \institution{University of Amsterdam}
  \city{Amsterdam}
  \country{Netherlands}
}
\email{yuhang.zheng0805@gmail.com}

\author{Xianzhe Fan}
\affiliation{
  \institution{Tsinghua University}
  \city{Beijing}
  \country{China}
}
\email{fxz21@mails.tsinghua.edu.cn}

\author{Bingbing Zhang}
\affiliation{
  \institution{School of Journalism and Mass Communication, University of Iowa}
  \city{Iowa City}
  \state{Iowa}
  \country{USA}
}
\email{bingbing-zhang@uiowa.edu}

\author{Zhicong Lu}
\affiliation{
  \institution{Department of Computer Science, George Mason University}
  \city{Fairfax}
  \state{Virginia}
  \country{USA}
}
\email{zlu6@gmu.edu}

\begin{abstract}

Previous research pays attention to how users strategically understand and consciously interact with algorithms but mainly focuses on an individual level, making it difficult to explore how users within communities could develop a collective understanding of algorithms and organize collective algorithmic actions. Through a two-year ethnography of online fan activities, this study investigates 43 core fans who always organize large-scale fans collective actions and their corresponding general fan groups. This study aims to reveal how these core fans mobilize millions of general fans through collective algorithmic actions. These core fans reported the rhetorical strategies used to persuade general fans, the steps taken to build a collective understanding of algorithms, and the collaborative processes that adapt collective actions across platforms and cultures. Our findings highlight the key factors that enable computer-supported collective algorithmic actions and extend collective action research into the large-scale domain targeting algorithms.
 
\end{abstract}



\maketitle

\section{Introduction}

In December 2020, Weibo, one of China’s largest social media platforms, created a separate trending section for entertainment after fan communities repeatedly manipulated the platform’s trending algorithm to promote celebrities and entertainment topics. This change aimed to prevent entertainment content from overshadowing more relevant public news and significant events on the main trending list ~\cite{xinlang2022entertaining, zhang2023push, guangming2023panentertainment}. Similarly, in 2023, South Korea's leading music streaming platform, Melon, modified its ranking algorithm to exclude muted plays, a tactic K-pop fans used to artificially boost their idols' rankings. In 2022, this manipulation generated an estimated \$33.2 million in revenue ~\cite{dredge2023streaming}. We define this phenomenon as "\textit{fan-driven collective algorithmic action}," wherein large-scale fans organize and mobilize to strategically alter algorithmic outputs on social media, aiming to enhance the visibility of their favorite idols. This collective effort encompasses a range of tactics that fans believe can influence algorithms, such as promoting specific topics to trend or engaging in repetitive muted video plays to elevate their idols' rankings ~\cite{zhang2023push, xiao2023we}.

Past HCI research has explored how various kinds of users strategically develop an understanding of platform algorithms and subsequently negotiate with these algorithms ~\cite{devito2017algorithms, cotter2019playing, karizat2021algorithmic, devos2022toward}. For instance, Cotter discussed how Instagram influencers develop an instrumental interaction of algorithms to increase their visibility by discerning the operational principles of these algorithms ~\cite{cotter2019playing}.  Yu et al. discovered that delivery workers formed micro-scale algorithmic communities by establishing private social media groups for mutual support. In these groups, they collaboratively developed a folk understanding of algorithms and shared insights into algorithmic functioning and resistance strategies ~\cite{yu2022emergence}. These studies have investigated how individuals or small groups of individuals make use of or resist algorithms and progressively develop their understanding of algorithms. However, at the collective level, particularly when considering larger groups of millions or even tens of millions, it remains unclear to HCI scholars how communities collectively understand algorithms and how collective actions towards algorithms could be developed and organized. \add{Understanding these large-scale actions is crucial due to their societal and technological implications. They challenge algorithmic integrity, expose gaps in platform governance, and reflect how users actively reshape algorithms to their advantage. Studying these behaviors offers insights for creating fairer, more transparent, and resilient platforms while contributing to discussions on algorithmic accountability and ethics.} 

In contrast to the academic community's lack of research on large-scale collective action targeting algorithms, such behaviors are on the rise and are having a significant impact on both platforms and society ~\cite{bonini2024algorithms, trere2024amplification}. Citizens are increasingly dissatisfied with their passive role in the face of algorithms and are seeking to enhance their influence through algorithmic activism ~\cite{velkova2021algorithmic,bonini2024algorithms}. For instance, political groups have previously engaged in Google bombing by utilizing predictions of Google's algorithms to develop strategies for designing websites, thereby influencing Google search outcomes ~\cite{gillespie2019algorithmically}. In 2018, activists manipulated Google’s algorithm and organized a large group of political allies on Reddit, massively upvoting Trump-related posts, which caused Google’s ranking algorithm to associate the term ‘idiot’ with his image ~\cite{haynes2018search, gillespie2019algorithmically}. \add{Thus, such large-scale collective actions, while empowering for communities, may also pose significant challenges. These actions can undermine platform integrity by distorting algorithmic outcomes, overcrowding spaces meant for public discourse, and eroding trust in platforms. For instance, fan-driven actions ~\cite{zhang2023push} can overshadow critical societal discussions, manipulate public perceptions, and promote artificial visibility at the expense of organic user engagement. Platforms must balance accommodating grassroots participation with maintaining the authenticity of their algorithmic outputs, which becomes increasingly difficult as users refine their manipulation tactics.}

As previous scholars of critical algorithm studies have pointed out, a significant trend in digital platforms is that more and more groups are using algorithms as tools for social movements and for gaining visibility~\cite{bonini2024algorithms, trere2024amplification, gillespie2019algorithmically}. These groups amplify their influence over algorithms and hijack platform algorithms to achieve their own goals~\cite{trere2024amplification}. However, there is a lack of research in academia on how these groups mobilize and organize large-scale actions targeting algorithms.

Our research uses large-scale (millions) fan-driven, computer-supported collective actions targeting algorithms (occurring across multiple countries and platforms) as a case study to expand the understanding of HCI and related scholars regarding such large-scale collective algorithmic actions. \add{By studying fan-driven algorithmic actions, we can better understand the mechanisms and motivations behind these behaviors and inform strategies to address their potentially harmful impacts. Our findings can help platforms develop governance models that are both resilient to manipulation and inclusive of diverse user voices. This research also contributes to ongoing discussions about the ethical implications of user-algorithm interactions, providing actionable insights for creating systems that are fairer, more transparent, and resistant to abuse. These contributions are particularly valuable for the CHI community, which focuses on the intersection of human behavior, technology design, and governance ~\cite{duvall2018blacklivesmatter, jung2012fan}.}

Specifically, in this paper, we aim to address the following research questions:

\textbf{RQ1:} What motivates fans to mobilize collective algorithmic actions?

\textbf{RQ2:} How do fans build a collective folk understanding of algorithms?

\textbf{RQ3:} How do fans organize and execute collective algorithmic actions?

To address these questions, we conducted a two-year ethnographic study ~\cite{harrison2018ethnography} of fan-driven collective algorithmic actions from 2021 to 2023. Previous research on fan communities has shown that collective algorithmic actions are often led by a small number of core fans. This demonstrates a hierarchical structure within fan communities, where core fans have the authority to mobilize large numbers of general fans for collective actions. In contrast, general fans typically occupy a more peripheral position regarding power within these communities ~\cite{zhang2023push, zheng2023play, wu2021fan}. 

Therefore, we observe 43 core fans and their corresponding general fan groups in collective fan algorithmic actions in a two-year ethnographic study. Additionally, we conducted one-on-one interviews and focus groups with the 43 core fans. On top of that, we interviewed 15 general fans to supplement their perspectives on collective algorithmic action. All the interviewed core fans have extensive experience organizing collective algorithmic actions and hold central roles in their communities, managing fan clubs with over 100,000 followers and maintaining personal social media accounts with more than 5,000 followers each. The general fans, with at least two years of participation, follow the directives of core fans in these actions but lack decision-making authority and influence over strategy, focusing instead on execution and support.

Throughout this longitudinal observation, we studied how fan communities, as large-scale collectives, develop, refine, and test their understanding of algorithms, converting this understanding into collective algorithmic actions. We also investigate how a small number of core fans persuade, guide, and coordinate millions of general fans to participate in these daily collective efforts to influence algorithms. 

Our work's key contributions to the HCI community are as follows: (1) Through a two-year ethnographic study, complemented by semi-structured interviews and focus groups, we conducted an in-depth analysis of how millions of fans orchestrate influential collective algorithmic actions. (2) We advanced folk theories of algorithms from an individual dimension to a collective understanding and explored how these folk theories evolve into well-defined, context-specific algorithmic action. We also extend computer-supported collective action studies to the field of algorithms. (3) We explored the potential for large-scale, computer-supported collective action by examining fan-driven algorithmic efforts capable of mobilizing millions of users. Our study, encompassing multiple fan communities, nations, and platforms, demonstrates not only the broad applicability of our theory to future large-scale collective actions but also potential strategies for overcoming large-scale collective action dilemmas.

\section{Related Work} 
\subsection{Folk Theories of Algorithm} \label{sec:folktheories}
Folk theories of algorithms are user-generated explanations that help people understand how algorithms work and influence them ~\cite{ytre2021folk, eslami2019user}. Users develop various folk understandings of how algorithmic systems operate and produce different outcomes, considering that these systems remain opaque to users ~\cite{karizat2021algorithmic, shen2021everyday, li2023participation}. Research has explored how users develop these theories through their direct interactions with algorithms ~\cite{cotter2019playing, eslami2019user}, algorithmic gossip~\cite{bishop2019managing, ticona2023platform, zhang2023push}, and social interactions within their communities ~\cite{mayworm2024content}. Through these means, users come to understand that social media information and interactions are driven by algorithmic systems and even learn some details about how these algorithms work. However, they still do not know the specific mechanisms behind the algorithms. This lack of full understanding leads them to develop their own folk theories for understanding their personal experiences with algorithms ~\cite{shen2021everyday, eslami2019user, li2023participation}. 

HCI studies have examined how users perceive various aspects of algorithms, such as how they affect user visibility on social media ~\cite{alvarado2018towards, duffy2023platform, cotter2019playing, devito2022transfeminine}, the processes behind content moderation ~\cite{mayworm2024content, moran2022folk}, how algorithms facilitate user matching on dating apps and how news is curated by algorithms ~\cite{ngo2022exploring}. These previous studies reveal how users build and refine their understanding of algorithms based on their day-to-day experiences and observations. In this process of testing and updating, users can develop algorithmic folk theories that are sufficiently intricate and flexible ~\cite{karizat2021algorithmic}. As they interact with algorithms and refine their understanding through ongoing experiences and observations, these folk theories become more nuanced. Users adjust their beliefs based on new information and changing interactions, allowing their folk theories to evolve through an iterative process in response to varying algorithmic behaviors and outcomes ~\cite{devito2021adaptive, zhang2023push}. 

Despite this, these folk theories of algorithms are often not entirely accurate because algorithms remain a "black box" to users, who typically lack detailed knowledge of the platform's algorithms or system design. However, these studies suggest that users have the potential to actively drive the testing and refinement of algorithms, gradually improving the reliability of their folk theories. This user-driven approach can enhance algorithmic transparency and understandability from the user's perspective ~\cite{shen2021everyday, eslami2019user}.

\subsection{Computer-Supported Collective Action}\label{sec:collectiveaction}

Olson defines collective action as efforts by a group to achieve common goals, such as lobbying, enhancing competitiveness, or engaging in protests ~\cite{olson1971logic}. This action is driven by collective interests, where benefits are shared and non-excludable, leading to the free-rider problem, where individuals benefit without contributing. Olson argues this issue is more severe in larger groups due to difficulties in organization, higher persuasion costs, and reduced individual incentives ~\cite{olson1971logic}. Consequently, Olson~\cite{olson1971logic} notes that collective action often results in "collective inaction," where groups fail to mobilize despite shared interests.

Effective collective action requires an organizational mechanism to coordinate and incentivize participants, such as providing incentives, establishing structures, or guiding through leadership ~\cite{olson1971logic}. Factors like group size, proximity, meeting frequency, uniformity, and organizing technologies also play crucial roles ~\cite{olson1971logic, elster1985making}. Elster ~\cite{elster1985making} further highlights collective consciousness as key to addressing the free-rider problem. This awareness connects participation to personal and collective identity, often fostered by experienced individuals and opinion leaders.

Recent HCI researchers have increasingly focused on computer-supported collective action due to its potential to generate significant attention on social media and create real-world impacts ~\cite{shaw2014computer, salehi2015we, bennett2013logic, matias2016going, wu2022reasonable, huang2015activists, yasseri2016political}. Salehi et al. ~\cite{salehi2015we} found that the web has the potential to enable distributed collectives to address shared issues. By examining the collective actions of over 100 Amazon Mechanical Turk workers, they explored the dynamics of computer-supported collective efforts. They introduced the concept of tactical publics, highlighting the transition from public deliberation to tactical actions and identifying stalling and friction as significant challenges in this process. Matias ~\cite{matias2016going} examined how 2,278 online community leaders on Reddit coordinated collective actions to influence platform operators. Their efforts led to millions of readers being restricted from accessing significant portions of Reddit and persuaded the company to negotiate to address their demands. 

Shaw et al. ~\cite{shaw2014computer} identified five key stages in computer-supported collective action. The first stage involves identifying a problem and finding others who share an interest in addressing it. The second stage focuses on generating, debating, and selecting viable solutions to the problem. In the third stage, participants coordinate and prepare for action, followed by the fourth stage, where the collective action is executed. Finally, the fifth stage emphasizes assessing, documenting, and following up on the outcomes of the action, ensuring that lessons are learned and future efforts are informed.

However, it is important to note that many collective actions on social media fail to achieve their goals; for instance, over 99\% of Change.org petitions aimed at improving privacy protections never reach "victory" status ~\cite{wu2022reasonable, huang2015activists}. To conclude, many of these efforts fail because they bypass key stages in the model proposed by Shaw et al. ~\cite{shaw2014computer}, such as effective issue framing, consensus-building, and the structured coordination of collective actions. Users also often lack the specific computational tools to collectively action. 

Previous HCI research has rarely addressed successful computer-supported collective actions. Even when it does, it seldom involves actions on the scale of millions of people ~\cite{salehi2015we,matias2016going}, particularly those specifically targeting algorithms. Our work uses large-scale, cross-platform, and cross-cultural fan algorithmic collective action as a case study to evaluate the iterative process of such computer-supported collective action development. We discuss how collectives develop a unified understanding of algorithms, translate this understanding into action, and effectively coordinate efforts across multiple platforms and cultures to address the organizational challenges faced in previous collective action. Additionally, we explore the resources and theoretical frameworks required for organizing computer-supported collective actions, focusing on those targeting algorithms.

\subsection{Transformative Fandom}\label{sec:fanalgorithm}

Previous HCI research has investigated fan communities' transformative potential, focusing on their communal and collective aspects. Wang et al.\cite{wang2024counting} explored how privacy theory extends from individuals to collectives by examining fan self-disclosure practices in online communities. They found that fans engage in value construction, trust-building, and personal disclosure, fostering a supportive collective environment. Park et al.\cite{park2021armed} studied how online fan communities effectively collaborate on causes beyond their original purpose. They revealed that fans use shared values, teamwork, and established infrastructure to achieve large-scale collaboration and build mutual trust. For instance, ARMY, BTS's fanbase, organized the \#MatchAMillion campaign in 2020, raising over one million USD for Black Lives Matter. This transformative aspect has made fans a key area of study in transformative community and self-organizational behavior ~\cite{zhang2023push, wang2024counting, jenkins2012textual, dym2018vulnerable, fiesler2016archive}.

Our case study focuses on Korean Popular Music (K-pop) and Chinese Popular Music (C-pop) idol fans, whose communities are closely linked with the idol industry ~\cite{jenkins2012textual}. Fans derive emotional satisfaction from idols ~\cite{yin2020emergent} and contribute to their success through online promotion and gifting ~\cite{yin2020emergent, stanfill2013they, yang2009all}. These fan communities are well-structured ~\cite{wu2021fan}. Wu ~\cite{wu2021fan} described how fan clubs, managing various fan activities, are central to this structure. In Wu's framework ~\cite{wu2021fan}, fan clubs and their departments are core fans, while general fans support idols based on their guidance. Zheng and Xiao ~\cite{zheng2023play} emphasized the core fans' authority, noting that general fans support them through actions like retweeting and positive commenting and engage in significant data labor under their direction. 

Recent research focuses on fans strategically engaging with platform algorithms and developing collective algorithmic actions to enhance their influence and idols’ visibility ~\cite{zhang2023push, he2023civic, zheng2023play, zhang2020east}. Yin ~\cite{yin2020emergent} found that fans willingly engage in unpaid data labor intended to influence algorithms and increase idol visibility. This includes basic activities like liking, commenting, and reposting, as well as promoting hashtags and countering negative comments ~\cite{yin2020emergent, zheng2023play, zhang2020east}. For example, Chinese fans have normalized data labor on platforms like Weibo, seeking to optimize their strategies based on algorithmic rules. Yin ~\cite{yin2020emergent} observed that collective emotions within fan communities strongly motivate this labor. Long-term data labor has shifted the meaning of fan practices on social media from consumption and interaction to focusing on data contribution for boosting idol visibility. Zhang et al. ~\cite{zhang2023push} examined how fans perceive the link between their algorithm manipulation activities and idol visibility, using platform reactions as a basis. They explored how fans integrate computational tools with manual labor and found that algorithmic imaginaries spread from core groups to the wider fan base through hierarchical and vertical processes. 

Although these studies have explored the transformative potential of fans and the phenomenon of fans developing collective actions targeting algorithms, research is still lacking on how these large-scale collective actions are organized within fan communities. While earlier studies have provided a theoretical foundation for collective algorithmic imaginaries among fans ~\cite{zhang2023push,zheng2023play,zhang2020east}, our work primarily contributes by examining how these collective algorithmic understandings undergo iterative development through collective efforts, extend to actionable levels, and become integrated into everyday practices across the community.

\section{Method}
% \subsection{Research Design}

In this research, we conducted an in-depth two-year ethnographic study from 2021 to 2023, focusing on interactions with 43 core fans from 12 famous fanclub in China. The study involved observing these 43 core fans, as well as tracking and analyzing the daily activities of their respective fan clubs and the broader general fan groups, all with their consent. As a supplement to the ethnographic observations, we conducted one-on-one interviews with all 43 core fans and organized three focus group discussions to delve deeper into their perspectives. Additionally, to broaden our understanding, we conducted one-on-one interviews with 15 general fans from these fan clubs to gather insights into their experiences and roles within the community.

\subsection{Ethnography Context: Define ``Core Fans'' and ``General Fans''}

In this study on fans' collective algorithmic actions, we define "core fans" as those who hold organizational power within the fan community, possess remarkable reputations, and directly influence the organization of collective fan actions. In contrast, "general fans" are those who participate in collective algorithmic actions but do not directly engage in core decision-making or shape the overall strategy of the actions; their primary role is to execute tasks as directed. 

\add{To be more detailed, in Chinese pop fan communities, interactions between core fans and general fans primarily occur online. Core fans, known for their dedication and critical influence within the fanbase, often go beyond virtual interactions to meet with other core fans in person. These in-person meetings are usually tied to organizing fan events, such as planning promotional campaigns, or hosting fan gatherings. For example, core fans may come together offline to design and execute support projects for their idols, like billboard advertisements. However, when it comes to general fans, their engagement with core fans tends to remain online. Core fans often take deliberate steps to maintain their privacy, avoiding the disclosure of their real identities to general fans. This trend has also been noted in prior research on Chinese fan cultures ~\cite{xiao2023we}, where it is highlighted that core fans carefully manage boundaries to preserve their influence and anonymity within the community. While some general fans might participate in offline events, they typically remain unaware of the true identities of the core fans who organize these activities.}

\add{Some core fans, according to our research, can even communicate directly with idols or their management teams. However, the precise nature of these relationships, such as the extent to which core fans influence idols’ decisions or public image, remains underexplored in current fandom studies. Core fans heavily depend on social media platforms to coordinate fan activities, disseminate information, and amplify their idols’ visibility. In some cases, core fans may gain access to privileged information through employees of these platforms, but this is neither a requirement for becoming a core fan nor a common trait among all core fans. Instead, the defining characteristics of core fans lie in their significant organizational skills and ability to rally large groups of fans within the community. These attributes grant core fans their elevated status and central role in fandom activities, regardless of whether they have insider access to information.}

\add{To clarify further, it is not the case that each core fan corresponds to a distinct and separate \textit{fan club}. Some of the core fans we interviewed belong to the same fan club. Our research involved a total of 12 fan clubs. However, the core fans we interviewed in the same fan club all held different positions to ensure the diversity of the research. At the request of our participants, we have not disclosed with which fan clubs individual participants are affiliated, as doing so could enable some readers to potentially infer the specific fan clubs and potentially identify the real identities of the participants.}

\subsection{Participating Role}
Inspired by previous HCI and fan studies ~\cite{thach2023key, zheng2023play, joshi2024subjectivity}, this research adopts an ethnographic method with a participant-observer approach. We reached out to 60 core fans, and ultimately, 43 of them agreed to participate in the study on the condition of complete anonymity. With their consent, two of our research team members immersed themselves in the fans' daily algorithm-related activities, which included internal discussions among core fans about algorithm action, designing a rhetorical language for persuading general fans, deciding which social media activities required collective algorithmic action, evaluating the effectiveness of these actions, and iterating on the processes. These researchers actively participated in these core fans' internal discussions.

\add{In fact, as researchers, we inevitably became part of the collective algorithmic action practices led by core fans, even shaping some aspects of these practices. For instance, during internal debates on how to optimize certain algorithmic practices, our feedback and observations, informed by academic perspectives, were occasionally incorporated into the core fans' plans. This dual role of observer and inadvertent participant underscores the complex dynamic between researchers and the communities they study in ethnographic research, particularly in contexts where participation is integral to understanding collective actions.}

Furthermore, these two researchers also engaged in the algorithmic practices typical of general fans, dedicating 1 to 2 hours daily, from Monday to Friday, to perform routine activities such as repetitive liking, reposting, and commenting, and adjusting specific strategies across different platforms. The platforms where our researchers conducted these algorithmic practices included YouTube, Weibo, X (formerly Twitter), Douyin, TikTok, Xiaohongshu, and other popular platforms where fan activities are active. \add{We did not participate in all 12 fan clubs' algorithmic practices every day. Instead, based on the core fan clubs' activity levels and relevance to our research questions, we selectively engaged with specific fan clubs on different days. } Additionally, they actively joined and observed the online communities of general fans within the fan clubs corresponding to these 43 core fans, participating in their discussions.

\add{We chose this participatory method as ethnographers because it provided a unique opportunity to gain in-depth insights into the nuanced, day-to-day dynamics of algorithmic collective actions that would be difficult to capture through interviews or passive observation alone. The participant-observer approach allowed us to understand the strategies, decision-making processes, and internal rhetoric in real time, which is crucial for examining the interplay between fan communities and algorithmic systems. This approach also fostered a level of trust and recognition from participants, enabling us to access sensitive and nuanced conversations that might otherwise remain hidden. This is particularly significant in the context of fan studies, where previous research has highlighted the vulnerability of fan communities and their guardedness towards outsiders ~\cite{brown2012ethnography}. These dynamics make ethnography not only a valuable but often the primary method for studying such communities ~\cite{brown2012ethnography}.}

\add{We acknowledge the need to reflect on how our involvement might have shaped the outcomes. By embedding ourselves within the community and actively participating in their practices, we were able to build rapport with participants, gaining their trust and, in turn, their willingness to share insights that are critical for understanding their algorithmic practices. This dual role of observer and participant aligns with the established practices in ethnographic fan research, where trust-building and immersive engagement are essential for generating authentic and comprehensive data.}

\subsection{Participants: Core Fans (N=43) and General Fans (N=15)}
The 43 Chinese core fans (CF1 to CF43) we interviewed all actively participated in organizing daily fan algorithmic actions. In addition, each organized over 30 specialized collective algorithmic actions targeting specific events, such as brand campaigns involving their idol, showcasing their extensive experience in coordinating such activities. The platforms impacted by the algorithmic actions discussed in these interviews included major fan activity platforms like YouTube, Weibo, X (formerly Twitter), Bilibili, Douyin, TikTok, Xiaohongshu, WeChat Channel and various music charts. Among the Chinese core fans we interviewed, 4 (CF1 to CF4) managed fan clubs with 100,000 to 200,000 followers on their social media accounts, 21 (CF5 to CF25) managed clubs with 200,000 to 500,000 followers, 11 (CF26 to CF36) managed clubs with 500,000 to 1,000,000 followers, and 7 (CF37 to CF43) managed clubs with more than 1,000,000 followers. In their view, following the idol might indicate an appreciation for the idol personally, but it does not necessarily imply active participation in algorithmic actions. In contrast, those who follow a fan club are much more likely to be highly motivated to engage in fan-driven algorithmic activities. Therefore, the data from these fan clubs can assist HCI researchers in understanding the scale of their organized algorithmic actions. These interviewed Chinese core fans could directly participate in and influence their fan clubs' decisions and judgments regarding algorithmic folk understanding and collective algorithmic actions. 

\add{Based on the core fans we interviewed, they believe that about 10\% of a fan club’s followers join in daily collective actions aimed at boosting their favorite celebrity’s visibility on social media. These daily actions are usually straightforward, such as liking, commenting, or sharing posts. However, when the tasks become more complicated, like organizing coordinated campaigns or using specific tactics to influence algorithms, only about 3\% to 5\% of followers participate. This number can vary depending on how loyal and united the general fans are. That said, during important events for their favorite celebrity, such as a new album release or a major award nomination, participation could go up to as much as 20\%. It’s important to emphasize that these numbers are not precise but rather the subjective estimates shared by the core fans we spoke to.} 

Additionally, all the core fans interviewed have their own social media account besides their fan club, while each had over 5,000 followers on social media platforms, with a monthly average readership exceeding 2 million, and their social media content primarily focused on their idol and fan community, thus affirming their core fans status (sufficient influence driven by their own fan identity).

Fifteen general fans (GF1 to GF15) were voluntarily recruited from the fan clubs of the previous 43 core fans. Each has over 5 years of fan experience and more than 2 years of participation in collective data actions. During the two years of our ethnographic study, they were actively engaged in algorithmic practices almost every day. They are distant from the decision-making groups of core fans and do not participate in the writing of algorithm tutorials or related discussions. Instead, they engage in the everyday algorithmic actions typical of general fans.

\subsection{Data Collection}
\add{The timeline of the data collection process spanned three phases: online and on-site observations conducted from 2021 to 2023, semi-structured interviews conducted in 2022, and focus group sessions held in early 2023. This phased approach allowed the researchers to iteratively refine their understanding of fan-led algorithmic practices, ensuring a comprehensive exploration of the phenomenon.} 

\subsubsection{Online and on-site observation}

We joined core fan groups on various social media platforms, such as WeChat, Weibo, and WhatsApp, where algorithms for organizing and coordinating actions are discussed, to understand how core fans perceive algorithms and how they extend this understanding to collective actions. Additionally, the two lead researchers participated in 22 offline gatherings of core fans to better grasp their perspectives and actions. We also joined the social media groups of general fans, focusing on discussions surrounding folk understanding of algorithms and the coordination of collective algorithmic actions. In these instances, the researchers were not passive participants; rather, they engaged in participatory observation as part of the fan community, fully immersing themselves in these algorithm-related fan activities ~\cite{zheng2023play}. 

\add{In most cases, we observed the internal discussions among core fans and the interactions between general fans on a daily basis across the 12 fan clubs. These discussions typically took place within different dedicated chat groups, reflecting the segmented nature of communication within fan communities. In accordance with the participants’ requests, we did not directly collect chat records using methods such as screenshots, but instead used secondary field notes.} 

This participatory observation approach follows an insider’s perspective, allowing the two lead authors to document and discuss the activities of core fans and the virtual or physical spaces they visited, thereby enabling interactive verification of data collection with other interviewees ~\cite{zheng2023play}. \add{Data collection during online observations included taking detailed field notes of the types of content shared, the language used in organizing algorithmic actions, and the dynamics of group interactions. During on-site observations at offline gatherings, the researchers recorded their observations in field journals immediately after the events to ensure accuracy and minimize recall bias. In both settings, the researchers adhered to ethical principles of informed consent and ensured that their presence did not disrupt the natural flow of fan activities.} These observations spanned from 2021 to 2023. 

\add{Throughout these two years, as the part of ethnography, we also engaged with Chinese core fans biweekly to gather their reflections on their daily collective algorithmic actions. These informal reflections were not structured as formal semi-structured interviews but were instead conducted in a conversational manner, resembling casual discussions.} \add{This provided a relaxed environment where participants felt comfortable sharing insights about both the internal strategies within the core fan groups and their influence on general fan groups. This approach aligns with ethnographic principles that emphasize immersion in the everyday lives of participants and the natural flow of information exchange.} \add{Regular interactions over an extended period enabled us to build trust with participants, providing a deeper understanding of the shifting dynamics within fan communities and the evolving nature of algorithmic practices.} \add{It also allowed us to document not just isolated actions but the iterative processes and ongoing adaptations that underpin successful collective actions.}

\subsubsection{Interviews with core fans and general fans}

Semi-structured interviews often effectively complement ethnographic observation data ~\cite{thach2023key,schensul1999essential}. Therefore, we conducted semi-structured interviews with the 43 core fans we observed and an additional 15 general fans from their fan clubs. These interviews were conducted online following the ethnographic observations to gather their views on collective algorithmic actions. All interviews were recorded and transcribed by the researchers. \add{The semi-structured interviews with core fans included open-ended questions such as: "What motivates you to organize algorithmic actions?" "How do you perceive the effectiveness of these actions?" and "Can you describe the process of designing persuasive messages for general fans?" For general fans, questions included: "What motivates you to participate in algorithmic actions?" "How do you perceive the instructions provided by core fans?" and "What challenges, if any, have you faced in engaging with these actions?" These interview questions were piloted with a small group of fans to ensure clarity and relevance.} 

The semi-structured interviews with core fans focused on their motivations for organizing collective algorithmic actions, challenges, the distinctions between different algorithmic actions, the iteration processes, and the reasoning behind these iterations. These interviews for core fans ranged from 1 to 3 hours, depending on the core fans' availability. For general fans, the semi-structured interviews explored their motivations for participating in collective algorithmic actions, their perceptions of the effectiveness of these actions, their views on the persuasive rhetoric and algorithmic action tutorials organized by core fans, and their assessment of the overall performance of the fan communities and fan clubs. These interviews for general fans ranged from 1 to 1.5 hours. 

\subsubsection{Focus groups with core fans}

After conducting observations and interviews, we held three focus group sessions with core fans, each with an average of five participants lasting approximately two and a half hours, to gather a collective perspective on algorithmic understanding and actions among core fans. Each focus group session includes five different participants. \add{The focus group sessions were guided by a researcher moderator, who posed open-ended prompts such as: "What are the key challenges in organizing collective algorithmic actions?" and "How do you as a group decide on strategies for engaging general fans?" The moderator ensured that all participants had the opportunity to contribute while maintaining the focus of the discussion. These sessions were recorded with participants’ consent, and transcripts were analyzed to identify themes related to group decision-making, shared challenges, and communal learning processes.} 

In HCI research, focus groups are a common complement to ethnography, enabling the exploration of group dynamics, collective reasoning, and the communal construction of knowledge. Unlike individual interviews, focus groups facilitate interactions among participants, bringing to light shared concerns, common strategies, and differing viewpoints, all crucial for understanding the collective behaviors that drive fan-led algorithmic practices. 

\subsection{Data Analysis}

\add{In this study, we employed a mixed-methods ethnographic approach, integrating field notes, interview transcripts, and focus group discussions to capture the multifaceted dynamics of our research context. Following established practices in ethnographic research, such as those outlined by Braun and Clarke in their thematic analysis framework ~\cite{braun2012thematic, braun2019reflecting}, we analyzed all data collectively without segregating it by source or method. This reflects the ethnographic tradition of treating diverse data as complementary pieces of a holistic puzzle, where multiple methods serve to triangulate findings and deepen understanding rather than isolating specific insights to a single methodological tool.}

Our data analysis followed a reflexive thematic analysis approach~\cite{braun2012thematic, braun2019reflecting}, in which two researchers collaboratively conducted qualitative coding. \add{The coding process began with both researchers independently coding a subset of data to familiarize themselves with the content and develop initial impressions. After this initial phase, the researchers convened to discuss their findings, consolidate preliminary codes, and create a unified codebook. This iterative process ensured alignment and consistency in coding while maintaining space for diverse interpretations. Separate codebooks were not created for the different data sources (observational data, interviews, focus groups), as our holistic approach treated these as interconnected data streams that contributed to a comprehensive understanding of the phenomenon. Instead, all data were coded using the same unified codebook, which evolved dynamically as additional insights emerged.}

\add{Through open coding, the data was systematically broken down into discrete elements, enabling the identification of underlying patterns, themes, and relationships. This step ensured that no potentially significant details were overlooked, as the researchers continuously revisited the data to refine their interpretations. Reflexive thematic analysis diverges from rigid, quantifiable measures of inter-coder reliability; instead, it emphasizes ongoing critical reflection and collaborative dialogue to ensure that diverse perspectives are considered throughout the analysis process.}

To synthesize and organize our insights, we used the affinity diagram method ~\cite{harboe2015real}, a widely recognized tool for clustering qualitative data into hierarchical themes. \add{The coding process resulted in the identification of a series of first-level, second-level, and third-level themes, each representing a progressively deeper abstraction of the data. This hierarchical organization not only helped to structure our findings but also allowed us to trace connections between granular observations and overarching theoretical concepts. The complete list of second- and third-level themes is provided in the ~\autoref{tab:higher-level-themes} in appendix.}

\add{By analyzing data holistically and integrating diverse methodological tools, we adhered to the ethnographic principle of contextual immersion. This approach emphasizes understanding the research phenomenon as a dynamic interplay of participant actions, cultural norms, and systemic influences. It also acknowledges the value of participant narratives, enabling us to co-construct meanings with our subjects rather than imposing rigid analytical frameworks ~\cite{braun2012thematic, braun2019reflecting}. This integrated and reflexive approach allowed us to derive rich, contextually grounded insights that form the foundation of our subsequent analysis.} We also present our research questions, corresponding findings, and their respective sections in \autoref{tab:rqs_findings} to enhance the clarity.

\subsection{\add{Positionality Statement}}

\add{We recognize that our personal experiences and backgrounds have shaped our research approach. Our team members bring diverse expertise in HCI, Communication and Sociology, with some research experience in folk theories of algorithms and fandom studies. Two authors have been actively involved in fan-driven algorithmic practices for over five years. One author, in particular, is an active core fan within her fan club, managing her own account in fandom that boasts nearly twenty thousand followers on Weibo and an average monthly readership of thirty million. She has extensive experience organizing fan-driven algorithmic actions across multiple countries, including China, South Korea, Malaysia, and Indonesia. Two authors have extensive experience in fan studies, while the other focus on algorithm research. In our research, the team members who are part of the fan community continuously emphasized the importance of adhering to community norms and ethical considerations. Meanwhile, the other researchers, serving as outsiders, provided a critical perspective. We regularly organized internal discussions among the researchers to ensure ethical considerations were upheld.}

\subsection{Ethics}
We required all participants to sign informed consent forms and consistently disclose our researcher identity throughout the study. Participants were given the option to withdraw from the research at any time. During the ethnographic observations, we adhered to ethical requirements by not directly collecting materials; the researchers compiled secondary field notes based on their observations for research purposes. In the interviews and focus group sessions, we sought participants' consent and encouraged them to ask any questions they had about the research. Throughout data analysis and manuscript preparation, we implemented strict anonymization procedures to protect participants' identities.

\begin{table*}[t]
    \caption{\add{Research Questions (RQs) and Key Findings}}
    \label{tab:rqs_findings}
    \centering
    \begin{tabular}{p{0.25\linewidth}|p{0.65\linewidth}}
        \toprule
        \textbf{Research Question (RQ)} & \textbf{Key Findings and Explanations} \\
        \midrule
        \textbf{RQ1:} What motivates fans to mobilize collective algorithmic actions? & \textbf{\autoref{rq1}:} Fans recognize the necessity of collective algorithmic action, believing their idol's visibility and success depend on collective efforts, often summarized as "all for our idol." \\
        \hline
        \textbf{RQ2:} How do fans build a collective folk understanding of algorithms? & \textbf{\autoref{rq2}:} Fans develop a collective folk understanding of algorithms, providing informal insights into how algorithms work across platforms and countries, forming the foundation for effective actions. \\
        \hline
        \textbf{RQ3:} How do fans organize and execute collective algorithmic actions? & \textbf{\autoref{rq3-a} and \autoref{rq3-b}:} Rational persuasion alone is insufficient. Effective actions rely on shared emotionality in fan communities, encapsulated in sentiments like "our idol only has us." Core fans must motivate general fans and use their folk understanding of algorithms to drive participation. \\
        \bottomrule
    \end{tabular}
\end{table*}

\section{Findings: “All for Our Idol”: Recognizing the Necessity of Collective Algorithmic Action (RQ1)}\label{rq1}
The first step in fan-driven collective algorithmic action is recognizing the necessity of such action. Core fans and general fans, when asked why fans engage in algorithmic action, unanimously see it as a key way to support their idols. They clearly understand that the behavior of fans on social media has a critical impact on the visibility of their idols, especially when influenced by recommendation algorithms. They have made a conscious connection between how fans engage with social media and the level of exposure their idols receive. By closely examining the various types of data generated by fan activities, they have been able to distinguish how different forms of social media engagement, such as likes, shares, comments, and other interactions affect the visibility and prominence of their idols within the algorithmically driven visibility of social media platforms. 

In the view of the core fans we interviewed, recognizing the impact of fan-platform interactions on their idols, as well as understanding the specific roles of different forms of interaction, is the first step in engaging in algorithmic action. This awareness makes them realize that taking algorithmic action is necessary and feasible. CF43 explains to us: \textit{"First, we need to understand that developing algorithms is beneficial for idols, right? And this benefit is quite useful; otherwise, we wouldn't be working so hard on this. Second, we must be clear about the practical impact of our approach; we can't just run around like headless chickens." }Next, we will explore how these fans perceive and interpret various fan behaviors on social media.

\subsection{The Importance of Data Performance in Idols' Commercial Collaborations}

Before organizing collective algorithmic actions, fans must first recognize that such actions are essential and necessary. Fans believe that an idol's data performance on social media crucially influences their commercial opportunities. CF38, a fan club leader, explains that brands don't just subjectively choose celebrities as ambassadors; objective data on social media is crucial. Brands work with data analysis firms and social media or e-commerce platforms for relevant data. This data meaningfully influences their choice of an idol. 

CF14, a fan club administrator, said that an idol partnering with a top brand gains not only notable monetary benefits but also promotional support from the brand. This includes invitations to events and shows sponsored by the brand. 

Certainly, impressive "data performance" is not limited to securing brand endorsements. CF42, a devoted fan with 15 years of fan experience, illustrated this point with one of her 2021 posts on Weibo, which stated that not only commercial brands but also reality shows and galas consider "data performance" when selecting their guests. While these platforms might not always seek assistance from professional data companies, given their less commercial nature, they still rely on intuitive and direct data metrics as iimportant factors in choosing celebrities as guests. Finally, this post garnered over 5,000 likes, 600 reposts, and 500 comments on Weibo, earning widespread recognition from the fan community. 

All core fans and interviewees in this study agree that "data performance" is vital for an idol's commercial appeal. CF32 explains, \textit{“Brands value an idol's data for gauging public visibility, reputation, and fan loyalty. Better data suggests more advantages for the idol."} She highlights fans' role, noting, \textit{“To improve our idol's chances, we focus on improving their data metrics.”}

\subsection{Perceived Relationship Between Fan-Driven Collective Algorithmic Action and Social Media Algorithms}

After recognizing the necessity of such collective actions, fans must also understand that their efforts, particularly large-scale collective actions, can indeed influence how algorithms perceive them. From the fans’ perspective, their activities on social media have the potential to affect and even manipulate these platforms’ algorithms.

Fans begin by distinguishing between data that does not require algorithmic intervention and data that does. This step establishes the necessity of taking action against the algorithms. CF29, a core fan responsible for organizing fan-driven data performance, outlined three main types of critical data: (1) \textit{monetary data}, (2) \textit{ranking-related data}, and (3) \textit{social media engagement data}.

\add{Monetary data includes sales figures related to idols, such as album sales, magazine sales, and endorsed product sales. This data often reflects the commercial influence of an idol’s core fan base, showcasing their purchasing power and the potential for brands to recoup costs through idol endorsements.} 

\add{Ranking-related data measures the time and effort fans are willing to invest in voting and boosting their idol’s rank. Fans may even act as “water armies”, a term for individuals hired (but in this case, unpaid) to manipulate online content to support brands. CF29 noted,\textit{ “When our idol is chosen as a brand ambassador, our data performance indirectly benefits the brand, as we engage in activities like sharing, liking, and commenting on specific posts or videos.”} She added, \textit{“The interaction data on social media directly correlates with the idol’s public exposure and reputation. Usually, better interaction data means more impactful promotion.” }CF33, an administrator of a fan data team, explained that in the realm of reality TV, an idol’s public visibility often serves as a predictor of a show’s potential ratings, making engagement data crucial in securing show invitations.}

The first two types of data, monetary contributions and ranking-related data, reflect the involvement level of an idol’s core fan base. Fans can easily improve these data metrics through repetitive actions like voting and spending money. 

However, the third type, social media engagement data, is more complex to manipulate because platforms use advanced algorithms to filter out artificial activity and focus on authentic engagement. To enhance performance in this area, fans must carefully devise strategies to influence the algorithms.

CF29 provided an example involving the ranking of comments on Weibo. According to CF29, Weibo uses a metric called “heat” to rank comments. Although the exact components of “heat” are not disclosed, it plays a critical role in determining comment rankings. Fans have deduced through observation that “heat” likely includes factors such as the number of likes, nested comments, account quality, and user interactions. They describe the dense layers of nested comments as akin to “buildings within buildings,” with hundreds or thousands of replies stacking up. Despite Weibo’s lack of transparency regarding the calculation of “heat,” fans strategically like and reply to favorable comments to increase their “heat” and boost their visibility. This practice, known as “controlling the comment section,” has drawn public criticism for affecting user experience, prompting Weibo to revise its “heat” rules multiple times. Nevertheless, fans continuously adapt their strategies to align with new algorithmic logic, creating a dynamic balance between the platform’s effort to ensure genuine engagement and fans’ efforts to promote their idols.

\add{In summary, the three main types of data, monetary data, ranking data, and social media engagement data, represent distinct dimensions of fan-driven data performance, reflecting varying levels of effort, strategy, and influence. Monetary Data demonstrates the purchasing power of an idol’s core fan base, as seen in sales of albums, magazines, and endorsed products, showcasing their direct commercial value to brands. Ranking Data highlights fans’ dedication through repetitive tasks like voting, improving idols’ positions in rankings to showcase their popularity to the public and investors. These rankings often lead to rewards like increased exposure or credibility from prestigious awards. Social Media Engagement Data focuses on influencing recommendation algorithms. Unlike ranking data, “heat” is a more abstract concept without a transparent calculation method, leaving fans to rely on trial and error to understand it. For example, fans may strategically like and comment on TikTok videos to increase their visibility, a tactic also commonly used by international K-pop fans. Algorithms on platforms like X trends, Billboard rankings, and iTunes rankings are similarly influenced by overseas fans’ practices. In addition to platform-driven recommendations, social media engagement metrics, such as likes, comments, and shares, have become essential indicators of an idol’s popularity and interaction levels. These metrics are often benchmarks for comparison among fan communities, further solidifying their importance in fan strategies and idols’ success metrics.}

\section{Findings: Fan-Driven Folk Theories: The Foundation for Guiding Collective Algorithmic Actions (RQ2)}\label{rq2}
The second step in collective algorithmic action requires a clear understanding of the different mechanisms by which algorithms operate on various platforms and even across different countries. Although this knowledge remains at a folk level, it must be detailed enough to effectively guide their actions. Next, we will demonstrate how core and general fans collaborate to develop a comprehensive understanding of platform algorithms.

CF26, a core fan within the Chinese fandom of a K-pop idol, emphasized that their primary role in collective algorithmic action is to mobilize general fans to engage in data labor. According to CF26, this task requires core fans to "decode social media algorithms" and guide general fans in manipulating specific data metrics, such as shares and comments, to improve the idol's overall data performance. Core fans must first understand how the algorithm functions and how to optimize content for recommendations. Only then can they make the process of algorithm manipulation understandable and executable, enabling even fans without technical expertise to participate effectively and follow the guidance provided.

\subsection{Fan-Driven Methods in Understanding Algorithms}

Before launching collective actions targeting algorithms, fans must have a method to thoroughly understand the algorithms, allowing them to perceive whether their actions can effectively influence them and adjust their strategies accordingly based on this assessment. Core fans use the term "decode" to describe the initial step in developing their understanding of algorithms. They see the algorithm as a black box; users are unaware of its exact operational logic and can only decode it through continuous trial and error. Eventually, this user-driven process creates a set of tutorials that can genuinely influence the algorithm's output in fans' perception.

CF26 explained that the process of algorithm decoding by core fans is largely based on folk empirical methods, as they lack direct support from the platform and cannot analyze it from a code or computational perspective. Their approach typically involves trial and error to develop the most effective strategies. 

To develop a clear understanding of the algorithm, initially, core fans often gather basic tutorials from other fan groups, as these groups also engage in decoding algorithms and instructing their fans. This is because all fan groups are actively decoding various social media algorithms to increase the visibility of their idols. To engage more general fans, these algorithm manipulation tutorials are always made public. As previously mentioned, there is already a consensus among core and general fans that various interactions on social media  impact their idols. Then, it is important to understand which specific interactions can most effectively leverage recommendation algorithms to maximize the visibility of their idols.

By observing patterns in algorithm behavior from various idol fanbases, core fans can formulate an initial strategy for their own group. For example, CF26 and her core fan colleagues examine tutorials from different idol fanbases, selecting and integrating them based on their experience. The core fan then guides general fans in improving the idol’s visibility through continual experimentation, eventually tailoring an algorithmic tutorial that best fits their fanbase. This is because fan groups are heterogeneous. The behavior and characteristics of fans from different idol groups can vary hugely, so core fans must carefully understand and adapt to the style of their corresponding general fan group. Moreover, since these are folk understandings, the algorithmic tutorials created by other fan groups are not necessarily effective. Core fans are often skeptical of tutorials from other groups and prefer to develop their own understanding through repeated experimentation and testing.

Improvements in data performance are often perceptible only to core fans, rather than being quantifiable through objective social media metrics. They observe whether the manipulations lead to higher comment rankings and visibility than other fanbases. Recently, CF26 got information from a fan working at Sina (Weibo company) about how to make a topic become a trending topic. When a topic related to an idol enters the trending section by the recommended algorithm, it gains visibility, making this achievement extremely important for fans. 

Within a single topic, there are often many posts, but how to intervene so that the algorithm recommends the topic and pushes it into Weibo's trending list, thereby gaining greater visibility, is a common concern for fans. CF26 mentioned that the new information from the Sina staff asks the fans to engage with the post displayed on the top of the topic, called “source of trend.” This means that interacting with the top posts within a topic, by giving them more retweets, likes, and comments, will make the topic more likely to enter the trending section rather than interacting with other, less prominent posts. When CF26 and her followers tried to make a new trend topic for their idol, they utilized this information to make a new tutorial, and engaged with the “source of trending.” After the topic successfully becomes a new trending topic, CF26 uploads a new post: \textit{“The new tutorial is so useful! It has been trending for about two hours!”} CF26 told us that she uses this period as a verification period to ensure the new tutorial makes sense.

\subsection{Expand Folk Understanding of Algorithms across Platforms in Different Nations and across Different Fan Bases}

If such collective action involves multiple countries, fans must be able to understand the differences across these countries. The essence of these fan algorithmic practices lies in a mindset centered on “algorithmic decoding” according to our participants, which guides fans’ collective actions across various platforms and countries. This means that, regardless of location, fans are essentially trying to understand the operational logic of the algorithm in order to manipulate it in reverse, increasing the visibility of content related to their idol. This approach is not confined to a single platform or nation; fans worldwide, particularly K-pop fans, apply this mindset to collectively influence data related to their idols on almost all platforms, often generating significant impact. For instance, on social media platforms like X (formerly Twitter), Instagram, and Weibo, fans might launch large-scale campaigns where they post the same hashtag en masse to boost its visibility or extensively like their idols’ posts to demonstrate popularity. On video platforms with recommendation algorithms, such as YouTube and BiliBili, fans organize targeted efforts to like their idols’ videos and repeatedly play them in specific ways to inflate view counts. Similarly, on short video platforms like TikTok, fans engage in activities such as liking, commenting, and fully watching their idols’ videos to enhance their recommendation metrics. Even on niche voting apps, fans often participate in repetitive and highly organized voting campaigns to support their idols. This fan- driven algorithmic practice is not restricted by platform or national boundaries. Instead, it reflects an internalized mindset within fan communities, driving organized collective actions that are applied broadly across the internet to amplify their idols’ presence and influence.

If this collective action involves different groups of fans, it is essential to understand the unique characteristics of these groups and leverage these traits to effectively organize them. Consequently, the interpretation of algorithms and the formulation of tutorials for algorithm manipulation by core fans vary between cultures and fanbases, and are largely based on empirical observations. 

In the context of the entertainment industry and social media, Chinese fans often employ a specific "nested comment (buildings within buildings)" strategy, which means comments within comments, a hierarchical commenting method used to manipulate data performance. CF26 highlights the unique dynamics between Chinese K-pop fans and fans of Chinese celebrities across social media platforms, noting: \textit{"The habits and scale of Chinese K-pop fans differ from those of fans of Chinese celebrities. They are smaller in scale and easier to manage."} In China, the "die-hard fan" status is a feature on Weibo that indicates a high level of interaction between a fan and a specific content bloggers. To obtain this status, a fan must interact with the bloggers for at least five days within the past 30 days. This status must be acquired separately for each blogger. The "nested comment" strategy in China is influenced by users' "die-hard fan" status and Weibo VIP membership. CF26 explains that if users meet both conditions, their fan community’s comments will gain higher visibility. However, if they lack "die-hard fan" status, this strategy may actually have a negative impact.

Furthermore, CF26 details that most general fans do not meet the criteria for either "die-hard fan" status or VIP membership. The "die-hard fan" status represents high-frequency interactions with a specific blogger, while obtaining VIP membership requires monetary contributions. General fans also find it difficult to maintain "die-hard fan" status across multiple bloggers simultaneously. As a result, the "nested comment" strategy could reduce visibility if it is employed by users who fail to meet these two requirements.

In this context, the term "blogger" here broadly refers to all Weibo users who post content, not just professional content creators or influencers. Essentially, any individual who publishes content on Weibo can be considered a "blogger." For general fans, maintaining "die-hard fan" status across multiple bloggers is challenging, as it requires consistent effort and resources. Consequently, the "nested comment" strategy intended to amplify visibility on social media may backfire if participants do not meet the "die-hard fan" and VIP requirements. This could lead to a decrease in the visibility of collective algorithmic actions, as these strategies heavily rely on the contributions of highly engaged and financially supportive fans.

CF43 agrees that for fans of Chinese celebrities, core fans often advise general fans against using the "nested comment" strategy. CF43 elaborates: \textit{"This is because Chinese celebrities have a large and uncontrollable fan base on Chinese social media. Directly instructing all fans to avoid using nested comments is a simpler and more efficient approach. When general fans, especially those who do not meet the aforementioned criteria, attempt to use this strategy, the platform may classify their actions as 'bot-like' behavior, considering it a disruption to social media order. Once flagged as 'bot activity,' fans’ comments will be downgraded and pushed to the bottom of the comment section, losing all visibility and rendering previous efforts futile."}

In contrast, Chinese K-pop fans, due to their smaller scale and more clustered activities, enable core fans to exert greater control over general fans, permitting only those with "die-hard fan" status and VIP membership to participate in "nested comments." The effectiveness of this strategy among Chinese K-pop fans highlights the different ways fan communities adapt to the same platform algorithms. Despite facing similar requirements for "die-hard fan" and VIP membership, fans of Chinese celebrities, being more difficult to manage, opt to ban all fans from using "nested comments," whereas K-pop fans can better coordinate their fanbase, allowing them to leverage the "nested comment" strategy.

CF11, who manages a data group in an idol's fan club, also highlighted how fans from different countries, even those of the same idol, employ distinct strategies on platforms like YouTube. CF11 noted that view counts of fancams on YouTube are crucial for K-pop idols, and K-pop fans worldwide, regardless of their country, strive to increase these counts. However, CF11 pointed out a stark contrast in techniques: while international fan organizations advise fans to remain logged in while viewing, Chinese fans do not recommend this approach. CF11 further explained their understanding of YouTube's view count algorithm: \textit{“YouTube’s algorithm limits the number of view counts to 3-5 plays per video per day from a single user. Users logged in can't contribute more than five views. Therefore, to exceed this limit, one should not log in.”} CF11 believes these different approaches stem from the varying user habits of the two fan bases, \textit{“Chinese fans are used to repeatedly engaging in data labor, often exceeding five views per day, hence our preference for not logging in. In contrast, the international fan base, predominantly from Europe and the U.S., tends to view 3-5 times daily. For them, ensuring each view is counted becomes vital, making logged-in viewing more appropriate.”} 

These differences between fan communities also suggest that if core fans do not develop a precise understanding of platform algorithms, their collective actions are unlikely to succeed. Without such knowledge, they cannot determine which behaviors will be effective or ineffective, and may even inadvertently trigger negative consequences.

\subsection{Expand Folk Understanding of Algorithm across Different Levels of Algorithm Transparency}

Since this is a collective action targeting algorithms, it must take into account the differences between the algorithms of various platforms, particularly in terms of transparency. Effectively leveraging platform transparency can enhance the efficiency of the action. Even when dealing with platforms with low transparency, fans must strive to understand the underlying algorithmic mechanisms. While deciphering algorithms can be intricate and challenging, as seen with platforms like Weibo, some platforms offer more transparency. CF39 mentioned that many platforms provide official explanations of their recommendation systems, simplifying the process for core fans to create tutorials. 

For example, BiliBili, a popular platform in China, officially states that the same video can be counted 3-5 times per day per person for play data. Although full playback isn't necessary for counting views, the completion rate in playing is a perceived important factor in the recommendation algorithm. This clarity gives fans straightforward tutorials for contributing to BiliBili's \textit{play count}. 

However, according to our participants, the criteria for BiliBili's \textit{trending list} are less clear for fans. Within the Chinese fandom, it was rumored that a video must achieve 100k plays within 24 hours, along with a specific number of likes, shares, comments, and other metrics, to feature on the trending list algorithmically. Yet, there are instances where videos have made it to the trending list with fewer than 100k plays or exceeded this threshold without being featured. CF39 expressed frustration over this ambiguity, stating, \textit{"We understand that activities like plays, likes, favorites, comments, and shares positively impact trending potential, but the exact mechanism is unclear. We cannot definitively influence the recommendation algorithm. Our only recourse is for fans to engage in data labor blindly and keep experimenting."} (CF39)

This scenario highlights the often opaque and unpredictable nature of social media algorithms, leaving even experienced core fans like CF39 to resort to trial-and-error methods.

We have described how core fans gradually develop a folk understanding of how algorithms operate on different platforms. It is important to note that while these understandings are not entirely conveyed to general fans, they serve as a crucial foundation for core fans to develop algorithm manipulation tutorials. All the core fans we interviewed acknowledged that without a clear understanding of platform algorithms, collective algorithmic action would be impossible. CF34, a core fan with over ten years of experience, explained to us, \textit{"Ten years ago, we lacked understanding of algorithms and the idea of collective algorithmic action. Fans didn't unite. Even after recognizing the importance of recommendation algorithms, it took time to learn how to coordinate effectively. Without clear direction, mobilizing millions of fans would have been impossible. Now, our actions are organized and disciplined, and fans know precisely what is expected of them, resulting in a observed impact."}

In short, having a systematic approach that allows for iterative refinement of communities' understanding of algorithms and the development of detailed algorithmic theories tailored to the characteristics of specific groups across different platforms and countries is a critical step in organizing collective algorithmic action.

\section{Findings: “Enable the Foolish to Learn, Motivate the Lazy to Act”: The Principles of Mobilizing Millions of Fans (RQ3)}\label{rq3-a}
Once core fans are motivated to organize collective fan actions and have gained a sufficiently clear understanding of the underlying mechanisms of algorithms at a folk level, the next challenge is how to effectively mobilize a larger number of general fans to participate in these collective actions. In essence, the folk decoding of algorithms represents merely the initial phase in the algorithmic practices of fans, aimed at enhancing their favorite idol's data performance. The predominant challenge for most core fans is convincing a wider fan base to participate in this collective action for their idols. 

CF37 challenged the previous Chinese media's portrayal of fans as a highly efficient, organized group that immediately responds to requests for collective algorithmic action. According to CF37, this depiction is misleading. CF37 explained that each act of data engagement, likes, shares, or comments, represents not just a digital footprint but also the genuine affection of a real person behind the screen. As CF37 noted, \textit{"We, all the fans, including core fans like myself, partake in this data labor purely out of love, without any financial or other incentives. Our sole aim is to elevate our idol's stature."}

However, because this work is unpaid, many fans hesitate to dedicate their time and energy to such tasks. CF37 pointed out that core fans must rely on persuasion, emphasizing both the importance of the data and the simplicity of their tutorials to encourage wider participation in these practices. By framing data engagement as an accessible and meaningful way to support the idol, they hope to motivate more fans to join the collective effort.

According to the 43 core fans we interviewed, to encourage an increasing number of general fans to participate in these algorithmic actions, they have undertaken two primary strategies:
\begin{itemize}
\item First, they focus on convincing general fans of the significance of fan-driven collective algorithmic actions. 
\item Second, they aim to create straightforward, user-friendly tutorials that are easy to understand and implement for general fans.
\end{itemize}

\subsection{Convincing Skeptical General Fans}

All core fans in our cases reported they heard skepticism from general fans, who often question the effectiveness of their data labor: \textit{"Is it really useful for me to do this data? Is the time and effort I spent really rewarding my idol accordingly?"} 

As GF4, a general fan with two years of experience whom we interviewed, put it, \textit{"Participating in algorithmic actions is really exhausting and complicated. I have to figure out what I need to do on each platform, and I spend a lot of time browsing posts and even playing videos, which drains my energy and ties up my devices. Sometimes, I don’t even know what my actions mean. Does my small effort as a single user really make a difference? And if I don’t do it, there will be other fans who will."} 

All the core fans we spoke to believe their primary task, before presenting any tutorial, is to convince other fans of the importance of participating as data labor. This kind of persuasion is key to reducing the free-rider effect among general fans and encouraging active participation in collective algorithmic actions. Core fans inform general fans and make numerous posts about the direct benefits that fan data labor can bring to the idols, often citing examples of other idols' success stories. 

For instance, core fans will argue that if their beloved idol's data performance is outstanding, similar benefits achieved by predecessors can be expected. If the benefits are not immediately evident from past cases, core fans will highlight the authoritative nature of the platform where they engage in data labor. They also emphasize the prominence of idols who are part of the same data performance competition, thereby crafting a narrative that suggests brands and TV programmers closely monitor this data. This way, core fans imply to the general fans that their idol could gain resource boosts through this data and their collective labor efforts.

\subsection{Making Algorithmic Manipulation Tutorials as Simple as Possible}

Moreover, CF16, a core fan who frequently creates algorithmic manipulation tutorials for their fan group, emphasized the importance of making these tutorials as simple as possible for general fans. She explained that this often requires creative approaches to fit seamlessly into fans' daily routines. 

For example, CF16 explained that she developed two tutorials for boosting video plays on Bilibili: one for the mobile app, which requires fans to manually click every 20 minutes, and another for the web version, which can loop the video continuously with no further input, making it much easier and less demanding. Furthermore, CF16 observed that it is more effective to ask fans to boost video plays in the morning when they are at work and can use the web version on their computers. In the evening, fans are more likely to be at home using the mobile app on their smartphones, which is more demanding. As a result, she strategically times her calls for engagement during morning hours to maximize participation.

There are many instances where the tools available for fan activities differ depending on the device or platform used. For example, an auto-tapper app for sending love taps during live idol streams is only available to Android users, while Apple users must manually tap their screens at intervals. As a result, this task is mostly handled by Android users, as it is more cumbersome for those with Apple devices. Similarly, a Weibo-based tool named "SHARE", which simplifies "controlling comments" with a keyword filtering function, is also limited to Android devices. This led to Android users taking the lead in managing and evaluating content related to their idols in a long period.

However, CF16, CF19, CF36, CF42, and CF43 noted that this changed in 2021 when the SHARE app was reportedly taken down due to infringement issues with Weibo, leveling the playing field between Android and Apple users. This shift allowed Apple users to become more involved in "controlling comments," reflecting the evolving landscape of fan algorithmic activities and the tools they use.

\add{It is worth noting that during the execution of algorithmic practices, fans occasionally employ tools to ease their workload, transforming algorithmic labor into a semi-automated process. Some of these tools are purpose-built apps designed specifically for fan activities, such as Xingfantuan (an app that enables one-click operations for certain data-related tasks on Weibo) and auto-clickers (which allow Android devices to repeatedly tap specific areas on the screen). However, in most cases, fans repurpose existing features to meet their needs and achieve semi-automation. For example, iPhone users might leverage iOS's built-in AssistiveTouch or gesture controls for semi-automated tasks, while PC users might combine browser auto-refresh features with YouTube’s autoplay function to repeatedly play videos automatically.} \add{In essence, semi-automation represents a mindset rather than reliance on any specific tool. It reflects a core fan tendency to adapt and streamline instructional practices into automated workflows, showcasing their strategic and resourceful approach to algorithmic interventions.}

The key to an effective algorithmic tutorial lies in its simplicity and understandability. The complexity of these tutorials is often tailored to the specific fanbase, with their loyalty and dedication playing a crucial role. As CF40, a fan who has been a core fan in various fan groups, explained, \textit{"Core fans for idols with highly devoted fanbases typically employ more complex but detailed tutorials. In contrast, core fans for actors whose fans are mostly casual supporters tend to opt for simpler tutorials, sacrificing some level of detail to attract more participation in data labor." }(CF40)

Although this observation is not backed by quantitative data, there is a consensus among all the core fans we interviewed. The less manual intervention a tutorial requires, the more likely fans will use it for data manipulation. If the general fan can engage in algorithmic action through a straightforward, automated setup on their phones or computers, their willingness to participate is generally high. Conversely, tasks requiring more cognitive engagement and physical effort are met with lower enthusiasm. 

\begin{figure*}[hbt!]
    \centering
    \includegraphics[width=0.8\linewidth]{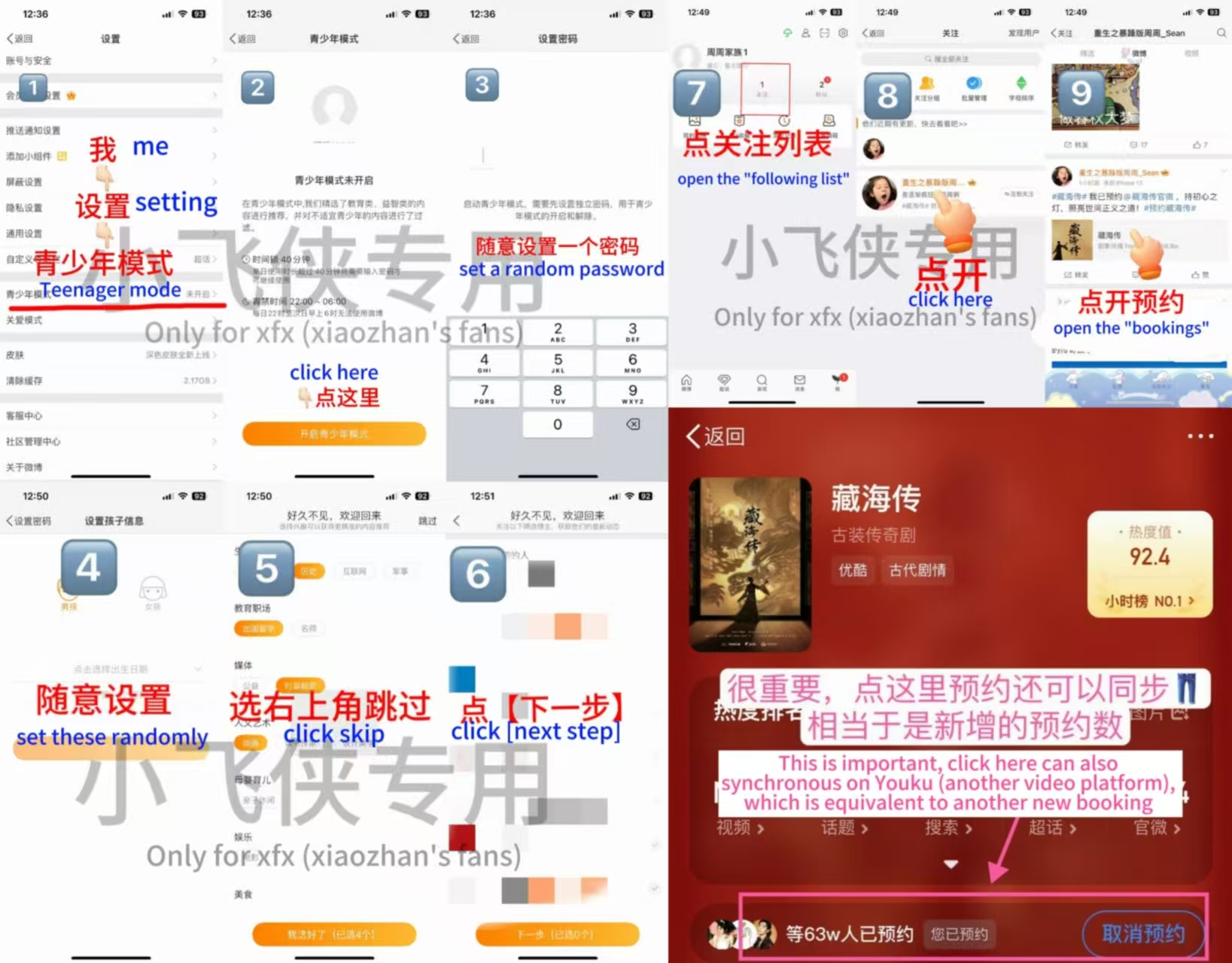}
    \caption{The figure shows a step-by-step tutorial created by fans of Xiao Zhan on how to make multiple reservations for his new drama Cang Hai Zhuan on Weibo's drama rating platform (similar to a Chinese version of IMDb). After making a reservation, the platform notifies users when the drama airs, so the number of reservations reflects public anticipation for the drama. Generally, each user can only make one reservation. Therefore, fans often register multiple accounts to make additional reservations.}
    \Description{The figure shows a tutorial created by fans of Xiao Zhan on how to make multiple reservations for his new drama Cang Hai Zhuan on Weibo's drama rating platform (similar to a Chinese version of IMDb). After making a reservation, the platform notifies users when the drama airs, so the number of reservations reflects public anticipation for the drama. Generally, each user can only make one reservation. Therefore, fans often register multiple accounts to make additional reservations.} 
    \label{fig:figure1}
\end{figure*}

\add{After summarizing and designing a streamlined and straightforward operational process, core fans typically compile it into a "tutorial" to facilitate understanding among fans. These tutorials generally come in three formats: text, images, and videos. Text-based tutorials are primarily used for very simple operations and are often shared informally between general fans for peer-to-peer teaching. Video tutorials, while capable of providing the most detailed instructions, are less convenient to disseminate and can deter some fans due to their length. Image-based tutorials are the most widely used and shared format, typically created by core fans or fan clubs and circulated by general fans. As CF3 noted, \textit{"Image tutorials are the clearest and most straightforward, making them the easiest for most fans to learn from. Only when some fans struggle to understand the image tutorials do we provide them with video tutorials for further guidance. However, most fans can grasp the instructions through image tutorials alone."}} CF42 concluded in one sentence, \textit{“In all, as core fans, we need to enable the foolish to learn while motivating the lazy to act.”}

\section{Findings: "Our Idol Only Has Us!” The Fan Algorithmic Emotionality of Collective Algorithmic Action (RQ3)}\label{rq3-b}
Rational persuasion alone is often insufficient to fully mobilize general fans. What supports large-scale mobilization is a powerful set of emotionality, which is widely recognized and continually reinforced within the fan communities. In their efforts to motivate general fans towards collective algorithmic actions, core fans often employ emotional elements alongside objective explanations of the data's importance. This strategy evokes emotional sympathy, encouraging fans to participate in collective algorithmic actions. Within the fanbase, this type of rhetoric is called "Guilt-Tripping Fans." 

"Guilt-Tripping Fans" involves conveying to fans that their idol is facing challenges and relies solely on their support, thus creating a sense of urgency and need. CF8, a core fan, describes this approach: \textit{"It is crucial for fans to understand the gravity of the situation, both rationally and emotionally. Our idols need us, not just in a figurative sense. Particularly in China, fans invest financially to fulfill a sense of obligation rather than merely seeking happiness. Therefore, fostering this sense of responsibility is key to compel fans to engage in data performance."} (CF8)

"Guilt-Tripping Fans" remains a contentious topic within the Chinese fandom community. This rhetoric is often seen as exploiting fans' emotions and coercing them into becoming data laborers. As GF8, a general member of the fanbase, observed, \textit{"In the fan culture,  any action that benefits our idol is regarded as a necessity."} She posits that this principle is powerful enough to influence fan behavior in various situations, akin to the effects of emotional coercion. 

CF39 elaborated on the strategic use of rhetoric in the context of "Guilt-Tripping Fans" and fans' algorithmic emotionality. CF39 said, \textit{"Effective 'Guilt-Tripping' rhetoric involves storytelling about our idol's past, present, and future."} CF39 further explained that the past rhetoric focuses on their challenges and perseverance, prompting fans to consider whether they want their idol's past efforts to be wasted due to their apathy. CF39 added, \textit{"The present rhetoric paints a picture of their current isolation and vulnerability."} CF39 emphasized that it highlights the idol's lack of support apart from fans, in an environment where company backing and capital often favor those with privileged connections, and external threats are constant. The message is that if even fans abandon them, their situation will become even more dire. CF39 continued, \textit{"The future rhetoric is about the potential outcomes of our actions or inactions."} CF39 elaborated that these scenarios are made as alarming as possible to compel everyone to act, aiming to prevent the idol from falling into unfavorable circumstances.

CF39 concluded that this narrative creates an atmosphere of guilt and regret, prompting fans to imagine the sadness they would feel if their idol were to suffer a tragic fate due to fan negligence. Once fans are convinced by this rhetoric, particularly when it is promoted by influential fans, they are likely to be swayed by the emotions embedded in collective algorithmic actions, effectively turning them into data laborers dedicated to these algorithmic efforts.

However, this "Guilt-Tripping Fans" rhetoric does not work for all general fans. CF37 points out that since the inflammatory rhetoric used by core fans is not mandatory, they cannot persuade every fan to act in the same way. \textit{"Of course, some general fans are dissatisfied with this, thinking, 'As long as I'm happy, I don't need to worry about how to help my idol,'"} CF37 adds, \textit{"But we actually have to accept the existence of such general fans, as they are only a small fraction of the fan base and do not affect the overall algorithmic emotionality within the fandom. This blending of algorithmic labor with emotionality towards loyalty and love for idols has already become an important part of fan culture and won't be disrupted by the minority who do not recognize it."} 

Furthermore, CF37 notes that in fandom. there is a comprehensive organizational strategy in place to ensure that most general fans involved in the fan community can perform algorithmic tasks. CF37 states, \textit{"General fans certainly have the option not to participate in these algorithmic actions or in the algorithm manipulation that we, as core fans, ask them to do. However, when it comes to distributing gifts considering idols from the fan club to general fans, such as priority access to idols’ concert tickets, these algorithmic labor activities are considered as key evaluation criteria."} Through this organizational strategy, the fan community disciplines its members, making fans who engage in algorithmic work the mainstream while those who refuse become the non-mainstream in fandom, unrecognized by the fan community and left on its periphery. In other words, these general fans, who do not participate in the fan-driven collective algorithmic actions, may continue to like their idols but find it challenging to integrate into the corresponding fan community.

"Guilt-Tripping Fans" resonates with previous research on algorithm-driven fan culture ~\cite{yin2020emergent}, which posits that data performance related to algorithms has become a pivotal object of fan sentiment and even a crucial factor in determining fans' self-identification and in securing recognition within the fan community. Ultimately, through the continuous emotional rhetoric, general fans cultivate an emotional stance within the algorithmic realm, believing that failure to engage in algorithmic labor or other supportive actions for their idols equates to a betrayal of their loyalty to the idol and a denial of their own identity as fans. Driven by the guilt instilled by "Guilt-Tripping" and their deep affection for their beloved celebrities, countless fans passionately partake in algorithmic labor.

\section{Discussion}\label{sec:discussion}

\subsection{From Individual Folk Theory to Collective Understanding and Action towards Algorithms}

Our findings contribute to the HCI literature by emphasizing the shift from individual folk theories to collective algorithmic understanding and action. \add{This research aligns with and extends existing theories on sociotechnical systems ~\cite{rogers2004new,rogers2012hci}, demonstrating how collective practices enrich our understanding of human-algorithm interaction and foster agency in digital ecosystems.} Specifically, we observe how minority groups decode algorithms, create actionable tutorials, and disseminate this knowledge among general users, thereby expanding the scope of folk theories of algorithms from understanding to actionable collective engagement.

This study corroborates the findings of previous critical algorithm studies research. Folk theories of algorithms ~\cite{devito2021adaptive, duffy2023platform, cotter2019playing, devito2022transfeminine} provide the theoretical foundation for the core fans to decode algorithms. Core fans continuously gather public information from platforms, professional folk understandings of algorithms from other fan groups, and gossip. As previous scholars have noted, these algorithmic insights from other groups form an external knowledge base that further refines their folk understanding of algorithms ~\cite{ticona2023platform, zhang2023push}. Core fans also observe changes in visibility following algorithm manipulation within their own fan community to refine their understanding of algorithms. The external knowledge base, combined with ongoing internal folk experiments, has led to an increasingly clear understanding of algorithms within the fan community.

\add{Our study extends the theoretical discourse by illustrating how grassroots experimentation within communities contributes to algorithmic literacy. This aligns with the broader HCI agenda of democratizing algorithmic agency, emphasizing the importance of user-driven innovation and shared knowledge in sociotechnical systems ~\cite{freeman2020pro}.}

Core fans cannot directly understand the platform's algorithmic mechanisms or access any quantitative data. Instead, their folk theories emerge from iterative observation and validation, a process that highlights the opacity of platform algorithms. \add{This iterative engagement reflects the "sensemaking loops" described in HCI ~\cite{aselmaa2017using,urquhart2024sense}, where users hypothesize, test, and refine their understanding of complex systems. Our findings underscore the importance of fostering algorithmic transparency to support user sensemaking and reduce reliance on speculative practices.} These iterations, while constrained by the black-box nature of algorithms, nonetheless provide practical foundations for mobilizing collective action within fan communities.

\add{Furthermore, our analysis reveals the centralization of algorithmic knowledge within fan communities, raising critical questions about power asymmetries in collaborative systems ~\cite{jhaver2021designing}. This underscores the dual nature of core fans ~\cite{zheng2023play,zhang2023push} as both enablers and gatekeepers of collective algorithmic practices, necessitating further HCI research on the implications of such dynamics for inclusivity and knowledge dissemination.}

\subsection{From Failed or Small-Scale Actions to Successful Large-Scale Computer-Supported Collective Actions}

Our research contributes to the theoretical understanding of large-scale, computer-supported collective actions in HCI. \add{By extending Shaw et al.'s five stages of collective action ~\cite{shaw2014computer}, we offer empirical evidence of how fan communities adapt these stages to engage with algorithmic systems. This demonstrates the relevance and flexibility of these theoretical frameworks in novel digital contexts, advancing HCI's understanding of algorithm-driven interactions.}

Instead of analyzing isolated actions, our study reveals the systemic processes through which core fans mobilize millions of participants across platforms and countries. This contribution lies in showcasing how users shape and influence algorithmic systems at scale. \add{By highlighting how fan communities integrate emotional motivation with practical strategies, our research expands the discourse on the interplay between emotionality and rationality in collective algorithmic practices ~\cite{etter2021activists, olson1971logic}, providing a new lens for studying user-driven digital interventions.}

Core fans identified a unifying need for algorithmic action to enhance their idol’s visibility, illustrating how shared intentionality drives large-scale mobilization. \add{This aligns with HCI theories on collaboration and resource alignment ~\cite{rozendaal2019objects,halabi2015exploration}, offering a concrete case of how shared goals facilitate effective collective action. Our study about fandom community also addresses ethical challenges, such as coercion and sustainability, contributing to the critical discourse on designing ethical collective practices in algorithmic contexts.}

\add{The iterative refinement of algorithmic practices by fan communities highlights adaptability as a key factor in successful collective actions. This finding resonates with HCI principles of iterative design and user feedback loops ~\cite{mirabdolah2023user,paton2021improving}, emphasizing the need for continuous evaluation and improvement in digital interventions. By linking user adaptability to collective success, our study enriches HCI's understanding of dynamic, user-driven processes in sociotechnical systems.}

Finally, our research underscores the replicability of collective algorithmic actions across diverse contexts. \add{We suggest that future HCI research explore how these practices can inform other domains, such as civic engagement or social movements, thereby expanding the scope of computer-supported cooperative work ~\cite{crivellaro2019infrastructuring}. By examining the interaction between collective agency and systemic constraints, our study offers a nuanced perspective on algorithmic interaction and its sociotechnical implications, broadening the horizons of HCI scholarship.}

\subsection{\add{From Emotional Engagement to Algorithmic Mastery: Rethinking Fan Practices in the Digital Age}}
\add{Fan studies have traditionally focused on fan identity, creativity, and cultural production ~\cite{duffett2013understanding,hills2003fan}. Our research expands this scope by highlighting how fans engage in algorithmic interventions to strategically enhance their idols’ visibility. By analyzing these collective practices, we introduce algorithmic systems as a critical site of fan activity, demonstrating how fans adapt their actions to the logic and constraints of digital platforms. This shift broadens the conceptual boundaries of fan studies to encompass the sociotechnical dimensions of fandom in the algorithmic era.}

\add{We contribute to fan theory by uncovering the interplay between emotional and rational dimensions of fan-driven algorithmic practices. While emotional investment in idols provides the primary motivation for participation, our findings reveal that fans leverage rational strategies, such as algorithm analysis, tactical voting, and coordinated social media interactions, to maximize their impact. This duality challenges traditional portrayals of fans as purely emotionally driven actors, positioning them instead as strategic agents capable of navigating complex systems. By bridging emotionality and rationality, our study offers a more nuanced understanding of fan behavior in the digital age.}

\add{Our research situates fan-driven algorithmic interventions within the broader discourse on fan labor. While prior studies have explored fans’ creative and promotional labor ~\cite{stanfill2014fandom}, we highlight a new dimension: algorithmic labor. This form of labor involves technical skills, collaborative coordination, and iterative experimentation, reflecting fans’ evolving roles as active participants in shaping algorithmic outputs. Furthermore, we demonstrate how collective fan actions align with and adapt theories of computer-supported cooperative work (CSCW) and collective action ~\cite{shaw2014computer}, offering a theoretical bridge between fan studies and HCI.}

\subsection{\add{Ethics in Future Platform Design and Governance}}

\add{The rise of large-scale, computer-supported collective actions highlights critical challenges and opportunities for platform design and governance. These collective actions, exemplified by fan-driven algorithmic interventions, reflect the growing agency of user communities in influencing algorithmic systems. While such practices can generate positive impacts, such as amplifying marginalized voices or fostering digital participation, they also pose ethical and operational concerns for platform integrity, fairness, and user welfare.}

\add{Large-scale algorithmic interventions demonstrate how collective user actions can shape platform dynamics, often bypassing intended algorithmic rules. From a governance perspective, platforms must navigate a fine line between fostering user engagement and mitigating manipulation. HCI design theories 
 ~\cite{zimmerman2014research,bardzell2012critical} emphasize the importance of incorporating diverse user perspectives into platform policies. Platforms could benefit from engaging user communities in the co-creation of governance frameworks, ensuring that collective actions align with ethical and operational goals while maintaining algorithmic integrity.}

\add{The ethical implications of collective algorithmic actions warrant closer attention. These actions often rely on emotionally charged mobilization strategies that can lead to coercion, burnout, or exclusion within communities. Additionally, large-scale manipulation of algorithms can overshadow organic content, distort public discourse, and exacerbate inequality in visibility. Addressing these risks requires a dual approach: tools that detect and monitor strategic (mis)behavior help identify manipulation in real-time, while value-sensitive design principles ensure that platforms proactively embed ethical safeguards to prevent exploitative practices. Drawing from HCI theories on value-sensitive design ~\cite{borning2012next}, platforms should integrate transparency mechanisms that clarify how engagement metrics are calculated, reducing misinformation about algorithmic processes. Moreover, participatory design approaches can involve users in shaping algorithmic policies, fostering collective accountability and mitigating harmful strategic behavior at its root.}

\add{A recurring issue in our findings is the difficulty users face in understanding algorithmic processes, not only due to their opacity but also because of their inherent complexity. This lack of clarity forces users to rely on speculative practices to influence outcomes. While transparency alone may not fully resolve this issue, platforms could adopt explainable AI approaches ~\cite{nazar2021systematic} to provide clearer insights into how algorithms function while maintaining system security. Additionally, promoting algorithmic literacy through educational tools and feedback systems could empower users to engage with platforms more responsibly and effectively, helping them navigate algorithmic decision-making beyond surface-level explanations.}

\add{The iterative and adaptive nature of fan-driven algorithmic actions underscores the need for platforms to design systems that are resilient to evolving user strategies. Drawing from HCI principles of iterative design and user feedback loops~\cite{mirabdolah2023user,paton2021improving}, platforms could implement adaptive algorithms that not only detect and respond to manipulation but also learn from user behaviors to improve their fairness and robustness. This dynamic approach can help balance the competing demands of user agency and platform governance.}

\add{Finally, it is essential to acknowledge that collective actions are not inherently good or bad; their impact depends on context and intent. Platforms should strive to distinguish between constructive uses of collective agency, such as civic engagement, and harmful manipulation, such as the distortion of public discourse. Building on theories of ethical technology design ~\cite{zimmerman2014research,bardzell2012critical}, platforms can prioritize interventions that amplify collective benefits while mitigating collective harms. For example, proactive moderation policies and inclusive community guidelines could help channel large-scale actions toward positive outcomes.}

\section{Limitations and Future Work}
This study has several limitations. First, it focuses on fan communities, which are renowned for their intense enthusiasm and high level of engagement with their idols, making them more likely to participate in collective actions than other user groups. Therefore, how to initiate large-scale collective actions among other user groups remains an open question. Second, the ethnographic research was conducted from 2021 to 2023, when fan collective actions had become routine and highly mature. We mainly focused on successful fan communities, making it difficult to observe moments of failure in collective actions and the factors contributing to such failures. We encourage future research to explore how computer-supported collective actions encounter failure, providing more case studies. Third, our study primarily examines fan collective actions related to algorithms, but fans also engage in collective actions in other areas. For example, previous scholars have examined American fans' collective support for democratic institutions or the Black Lives Matter movement ~\cite{duvall2018blacklivesmatter, jung2012fan}. We also encourage future scholars to investigate fan behavior in other forms of collective action.

\section{Conclusion}

In our study, we examine how millions of fans form effective, self-organized, computer-supported collective actions targeting platform algorithms. We highlight a distinct group of fans motivated by their affection for an idol and the aspiration to enhance the idol's success, which forms the foundation of their algorithmic emotionality. This unique emotionality framework in algorithmic engagement enables both the dedicated fans, known as core fans and the general fanbase to fully participate in algorithmic work. Core fans play a pivotal role in deciphering algorithms, creating comprehensive manipulation tutorials, and motivating general fans to partake in this algorithm labor zealously. To address the reluctance of most fans to invest substantial time and effort, core fans employ specific strategies: they leverage the emotional rhetoric within fan communities to continuously reinforce the algorithmic emotional stance among fans. Furthermore, core fans work to simplify and automate the process of algorithm manipulation, thus making it easier for general fans to participate in this labor-intensive activity.

\bibliographystyle{ACM-Reference-Format}
\bibliography{sample-base.bib}

\appendix
\section{Higher-level Themes}

We provide a summary of the higher-level themes we identified in Table~\ref{tab:higher-level-themes}. These themes broadly correspond to the section headers in our results sections. 

\begin{table*}[h!]
\centering
\caption{The four third-level themes and eighteen second-level themes we identified through data analysis.}
\label{tab:higher-level-themes}
\scriptsize
\begin{tabularx}{\textwidth}{p{0.2\textwidth} p{0.2\textwidth} X}
\toprule
\textbf{Third-level themes} & \textbf{Second-level themes} & \textbf{Explanation} \\
\midrule

Recognizing the Necessity of Fan-Driven Collective Algorithmic Action & The Impact of Social Media Action on Idol Visibility & Fans understand that their social media actions, such as liking, sharing, and commenting, directly influence their idol's exposure and visibility.  \\
 & The Importance of Data Performance in Commercial Collaborations & Fans realize that an idol's "data performance" plays a decisive role in brand selection and commercial collaborations. Understanding that enhancing data performance through collective actions can help their idol secure more business opportunities makes these fan-driven actions necessary and meaningful. \\
 & Relationship Between Fan-Driven Collective Algorithmic Action and Social Media Algorithms & Through practice, fans have discovered that their strategies (such as "controlling comments") can influence social media algorithmic outcomes, thereby increasing their idol's visibility. This understanding of how algorithms function, coupled with the ongoing adjustment of their strategies, further reinforces the perceived necessity of collective algorithmic action. \\
 & Competitive Arenas on Digital Platforms & Fans understand that actions on various digital platforms (e.g., trending topics, popularity rankings) are essential battlegrounds for increasing their idol's visibility. Their activities are not limited to one platform but span multiple digital domains, extending their actions into different levels and emphasizing the need for various collective actions across platforms. \\

\midrule

The Theoretical Foundation for Guiding Collective Algorithmic Actions & Empirical Methods in Decoding Algorithms & Core fans use empirical methods, such as trial and error, to decode social media algorithms. They often gather basic tutorials from other fan groups, adapting them based on their own experience and experimentation to formulate strategies tailored to their fanbase. \\
 & Adapting Strategies to Different Cultural and Social Media Contexts & Core fans recognize the importance of adapting strategies according to cultural differences and platform-specific policies. For instance, they develop distinct approaches for manipulating data on Korean versus Chinese social media platforms. \\
 & Platform-Specific Differences in Algorithm Manipulation & Different social media platforms, like YouTube, Weibo, and BiliBili, have unique algorithms that require tailored manipulation strategies. Core fans develop specialized methods based on the specific criteria and constraints of each platform. \\
 & Importance of Algorithmic Knowledge for Effective Collective Action & Core fans emphasize that a clear understanding of platform algorithms is essential for coordinating large-scale, collective algorithmic actions. Without this foundational knowledge, it would be impossible to organize and direct the efforts of millions of fans effectively. \\
 & Challenges in Manipulating Opaque and Unpredictable Algorithms & Despite efforts to decode platform algorithms, fans often face challenges due to the opaque and unpredictable nature of social media algorithms, leading them to rely on trial-and-error methods. \\
 & Evolution of Collective Algorithmic Actions in Fan Communities & The development of collective algorithmic actions has evolved over time, moving from a lack of understanding to a more organized and disciplined approach, where general fans know what is expected of them, resulting in more impactful actions. \\

 \midrule

The Principles of Mobilizing Millions of Fans & The Challenge of Mobilizing General Fans for Collective Actions & While core fans are motivated to organize collective actions, the real challenge lies in engaging a larger number of general fans, who may be less committed or lack understanding of the importance of these actions, to actively participate and contribute to the idol's data performance efforts. \\
 & Overcoming Skepticism and Reducing the Free-Rider Effect & Core fans face skepticism from general fans who question the effectiveness of their data labor. To overcome this, core fans emphasize the significance of collective algorithmic actions and provide examples of successful outcomes to encourage participation. Core fans also develop easy-to-follow tutorials for general fans to facilitate their engagement in algorithmic actions. These tutorials are designed to be integrated into the daily routines of fans, minimizing the effort required and increasing participation. \\
 & Tailoring Tutorials to Device and Platform Accessibility & Core fans create different versions of tutorials based on the specific devices and platforms available to fans, such as mobile apps versus web browsers, to optimize participation across diverse user groups. \\
 & Balancing Complexity and Participation & The complexity of algorithmic tutorials is tailored to the dedication level of the fanbase. Tutorials with higher levels of automation tend to have higher participation rates, while more complex tasks requiring manual intervention see lower enthusiasm from fans. \\
 & Challenges in Manipulating Opaque and Unpredictable Algorithms & Despite efforts to decode platform algorithms, fans often face challenges due to the opaque and unpredictable nature of social media algorithms, leading them to rely on trial-and-error methods. \\
 & The Principle of Effective Fan Mobilization: Simplicity and Motivation & Core fans follow the principle of making learning accessible for all and motivating even the least active fans to participate. The goal is to create tutorials that are simple, automated, and easy to use, thereby maximizing participation in data labor. \\

   \midrule

The Fan Algorithmic Emotionality of Collective Algorithmic Action & The Limitations of Rational Persuasion in Mobilizing Fans & Rational arguments alone are often insufficient to motivate general fans to participate in collective algorithmic actions. \\
 & The Strategy of Guilt-Tripping Fans & Core fans use the tactic of "Guilt-Tripping Fans" to create a sense of urgency and responsibility, thereby encouraging participation in data labor. \\
 & Algorithmic Emotionality in Fan Communities & The use of guilt-tripping within the fanbase is seen as a form of algorithmic emotionality, where fans are made to feel obligated to contribute to data labor as a duty to support their idol. \\
 
\bottomrule
\end{tabularx}
\end{table*}

\end{document}